\documentclass{ieeeaccess}
\usepackage{cite}
\usepackage{amsmath,amssymb,amsfonts}
\usepackage{algorithmic}
\usepackage{graphicx}
\usepackage{textcomp}

\usepackage{caption}
\usepackage{amsmath}
\usepackage{graphicx}  
\usepackage{booktabs}  
\usepackage{array}     
\usepackage{caption}   
\usepackage{multirow}
\usepackage{amsmath}
\usepackage{amssymb}
\usepackage[table]{xcolor}
\usepackage{url}
\usepackage{xcolor}

\usepackage{bm}
\makeatletter
\AtBeginDocument{\DeclareMathVersion{bold}
\SetSymbolFont{operators}{bold}{T1}{times}{b}{n}
\SetSymbolFont{NewLetters}{bold}{T1}{times}{b}{it}
\SetMathAlphabet{\mathrm}{bold}{T1}{times}{b}{n}
\SetMathAlphabet{\mathit}{bold}{T1}{times}{b}{it}
\SetMathAlphabet{\mathbf}{bold}{T1}{times}{b}{n}
\SetMathAlphabet{\mathtt}{bold}{OT1}{pcr}{b}{n}
\SetSymbolFont{symbols}{bold}{OMS}{cmsy}{b}{n}
\renewcommand\boldmath{\@nomath\boldmath\mathversion{bold}}}
\makeatother

\def\BibTeX{{\rm B\kern-.05em{\sc i\kern-.025em b}\kern-.08em
    T\kern-.1667em\lower.7ex\hbox{E}\kern-.125emX}}

\begin{document}
\history{Date of publication xxxx 00, 0000, date of current version March 25, 2025.}
\doi{10.1109/ACCESS.2024.0429000}





\title{Lateral Phishing with Large Language Models: A Large Organization Comparative Study}
\author{\uppercase{Mazal Bethany}\authorrefmark{1,3,*}, 
\uppercase{Athanasios Galiopoulos}\authorrefmark{1,*},
\uppercase{Emet Bethany}\authorrefmark{2,3},
\uppercase{Mohammad Bahrami Karkevandi}\authorrefmark{2,3},
\uppercase{Nicole Beebe}\authorrefmark{1},
\uppercase{Nishant Vishwamitra}\authorrefmark{1}, and
\uppercase{Peyman Najafirad}\authorrefmark{1,2,3}}

\address[1]{Department of Information Systems and Cyber Security, University of Texas at San Antonio, San Antonio, TX 78249 USA}

\address[2]{Department of Computer Science, University of Texas at San Antonio, San Antonio, TX 78249 USA}

\address[3]{Secure AI and Autonomy Lab, San Antonio, TX 78249 USA}

\address[*]{Equal Contribution}

\tfootnote{This material is based on research sponsored by the Department of Homeland Security (DHS), United States Secret Service, National Computer Forensics Institute (NCFI) via contract number 70US0920D70090004.}

\markboth
{M. Bethany \headeretal: Lateral Phishing with Large Language Models: A Large Organization Comparative Study}
{M. Bethany \headeretal: Lateral Phishing with Large Language Models: A Large Organization Comparative Study}

\corresp{Corresponding author: Peyman Najafirad (e-mail: peyman.najafirad@utsa.edu).}

\begin{abstract}
\label{abstract}

The emergence of Large Language Models (LLMs) has heightened the threat of phishing emails by enabling the generation of highly targeted, personalized, and automated attacks. Traditionally, many phishing emails have been characterized by typos, errors, and poor language. These errors can be mitigated by LLMs, potentially lowering the barrier for attackers. Despite this, there is a lack of large-scale studies comparing the effectiveness of LLM-generated lateral phishing emails to those crafted by humans.
Current literature does not adequately address the comparative effectiveness of LLM and human-generated lateral phishing emails in a real-world, large-scale organizational setting, especially considering the potential for LLMs to generate more convincing and error-free phishing content.
To address this gap, we conducted a pioneering study within a large university, targeting its workforce of approximately 9,000 individuals including faculty, staff, administrators, and student workers. Our results indicate that LLM-generated lateral phishing emails are as effective as those written by communications professionals, emphasizing the critical threat posed by LLMs in leading phishing campaigns. We break down the results of the overall phishing experiment, comparing vulnerability between departments and job roles. Furthermore, to gather qualitative data, we administered a detailed questionnaire, revealing insights into the reasons and motivations behind vulnerable employee’s actions.
This study contributes to the understanding of cyber security threats in educational institutions and provides a comprehensive comparison of LLM and human-generated phishing emails' effectiveness, considering the potential for LLMs to generate more convincing content. The findings highlight the need for enhanced user education and system defenses to mitigate the growing threat of AI-powered phishing attacks.

\end{abstract}


\titlepgskip=-21pt

\maketitle

\section{Introduction}
\label{introduction}

The rapid advancement of internet technologies has significantly heightened security concerns, especially in the realm of phishing, a critical form of social engineering that deceitfully extracts sensitive information \cite{gupta2016literature}, \cite{aldawood2019academic}, \cite{hadnagy2010social}, \cite{aldawood2018educating}. Attackers masquerade as legitimate entities to gain unauthorized access to accounts and acquire confidential data \cite{das2019sok}. The education sector in particular has been heavily targeted \cite{zscaler2023phishing}, and in 2024 continues to be the most heavily targeted sector \cite{checkpoint_research_2024}. Lateral phishing, which exploits both technical and human vulnerabilities by crafting personalized communications, has become a sophisticated and deceptive attack method \cite{bhadane2019detecting}. While knowledgeable users may not be easily deceived by less sophisticated attackers \cite{akhawe2013alice}, whose poor grammar and obvious inconsistencies give them away, the advent of Large Language Models (LLMs) like OpenAI's ChatGPT has complicated the landscape. By producing remarkably human-like text \cite{touvron2023llama}, \cite{zhao2023more}, \cite{herbold2023large}, these LLMs can empower even less advanced attackers, significantly broadening their capabilities to mount more sophisticated and convincing phishing attacks. Additionally, the use of AI in phishing has been noted to enhance the frequency and sophistication of social engineering attacks \cite{DarktraceEmailSecurity}.

This escalating threat landscape, particularly with the involvement of LLMs in phishing, necessitates innovative studies to understand this threat. There are three main problems regarding the existing literature with respect to this new threat. First, existing studies on lateral phishing attacks do not investigate the integration of LLMs for large-scale attacks targeting entire organizations. Previous works that analyze phishing attacks at scale, provide valuable insight into the tactics and techniques as well as mitigation strategies towards phishing threats \cite{ho2019detecting, lain2022phishing, oest2020sunrise}, but do not consider the role that LLMs could play in this threat. However, conducting such studies demands a real-world environment that is representative of a large-scale organizational structure. Second, while some recent work has been done towards the automated detection of LLM generated phishing emails \cite{roy2023chatbots}, there has yet to be work to quantify users' vulnerability to such emails on a large scale. Third, there's a notable gap in understanding the reasons why individuals might interact with lateral phishing emails, as there is a lack of studies directly asking those who fell for such attacks about their motivations and thought processes.

In this research, we aim to answer the following research questions: 1.) How does the effectiveness of lateral phishing emails generated by Large Language Models compare to that of those crafted by humans within an educational organization? 2.) Which job roles and department affiliations exhibit the highest risk of falling victim to phishing emails, as evidenced by elevated click-through rates, and what patterns can be discerned to identify these high-risk groups? 3.) What are the underlying reasons for individuals to engage with lateral phishing emails, as revealed through questionnaires with those who have clicked on them? Our investigation, situated within a large public university, first delves into the specific cyber threats faced by the institution. Investigating the history of security incidents within the organization and through discussions with the cyber operations team, we find that phishing campaigns are the most common security incident, with a 15\% increase in just the past year and that LLMs may be to blame. Informed by these observations, we conducted a large scale study across roughly 9000 employees where we tested a handful of different phishing email templates. The existing phishing simulation infrastructure of the university's cyber operations team supports the logging of information of how users interact with simulated phishing emails, collecting information such as if they open the email, click the phishing link or input login credentials. Furthermore, we conduct a questionnaire on those who input login credentials in the phishing experiment to understand why users interact with the phishing emails, in order to understand some of the key decision-making factors that lead individuals to fall prey to these sophisticated cyber threats, thereby informing more effective user education and system defenses. Our study shows that LLM-generated lateral phishing emails are as effective as, and in some cases even more effective than, those crafted by human communications professionals. Notably, in one of the most effective lateral phishing scenarios, where the sender is a compromised account of a supervisor, LLM-generated emails performed comparably to human-crafted ones, exploiting trust relationships and power dynamics within the organization. Furthermore, LLM-generated emails significantly outperformed human-crafted emails in the timely phishing scenario, capitalizing on current events and institutional context. Our analysis also identified that student roles exhibited the highest vulnerability to phishing attempts, with the highest percentages of both links clicked and data entered compared to other job roles. Additionally, insights gathered from a questionnaire administered to employees who entered data in the simulated phishing emails revealed that the sender's identity, perceived relevance of the email, and the sense of urgency were key factors influencing recipients' decision to interact with the phishing content.

Ultimately, this research provides the following key contributions. 
\begin{itemize}
\item We undertake a thorough study to show the state of cyber security threats at a large public university. Alongside this we show how insights from the organization's cyber operations team drive our investigation into LLM generated phishing attacks.
\item We conduct experiments across roughly 9000 employees at a large public university to compare how they interact with LLM written phishing emails versus human written phishing emails.
\item We administered a detailed questionnaire to 34 employees who entered data into the simulated phishing emails, aiming to understand the reasons behind their actions. This allowed us to gather insights, which can inform future strategies for phishing mitigation within the organization.
\end{itemize}

\section{Related Work}
\label{background}



Phishing, a widely recognized cybersecurity threat, has been extensively researched due to its prevalence in targeting both individuals and organizations \cite{shahbaznezhad2021employees, khonji2013phishing}. Characterized by its broad and indiscriminate approach, phishing attacks often aim to deceive a large audience through misleading emails or messages to illicitly acquire sensitive data \cite{alkhalil2021phishing}. The success of these attacks is largely attributed to the use of persuasive and deceptive techniques in crafting the content, which are designed to exploit human vulnerabilities \cite{fatima2019persuasive}. Other studies instead investigate the differences in how some people better manage phishing emails in their inbox \cite{pattinson2012some}. Transitioning from this broader perspective, spearphishing represents a more refined form of phishing. Distinguished by its highly targeted nature, spearphishing attacks are not only meticulously crafted but are also deeply personalized, targeting specific individuals or entities within organizations. These attacks often involve elaborate social engineering strategies, leveraging detailed knowledge about the target to increase their efficacy \cite{rajivan2018creative}. Recent studies indicate that spearphishing is not only more successful than general phishing but also more challenging to detect due to its tailored approach and exploitation of specific individual or organizational traits \cite{bhadane2019detecting}. The evolution of spearphishing tactics, including the use of sophisticated psychological manipulation and the exploitation of social networks, has significantly raised the stakes in the realm of cybersecurity threats \cite{parmar2012protecting}.

In recent years, the landscape of phishing attacks has become increasingly sophisticated and targeted. According to the Check Point Software Technologies Ltd.'s Brand Phishing Report for Q3 2023, Walmart emerged as the most impersonated brand in phishing scams, with Microsoft, Wells Fargo, and Google also being frequently targeted \cite{CheckPoint2023}. This trend indicates a strategic shift in attackers' preferences towards high-profile, globally recognized brands. Complementing this, Vade Secure's Q3 2023 Phishing and Malware Report indicates a significant uptick in phishing activities, with a 104\% increase in phishing URLs targeting Facebook and a staggering 973\% increase in phishing URLs aimed at Bank of America, illustrating the financial sector's increasing vulnerability \cite{VadeSecure2023}. Further, Zscaler ThreatLabz's 2023 Phishing Report observed a 47.2\% surge in phishing attacks in 2022, with the education sector being particularly targeted, underscoring the diversification in industries at risk \cite{zscaler2023phishing}. Finally, Barracuda Networks, Inc.'s Spear Phishing: Top Threats and Trends Report 2022 delves into the nuanced tactics of spear phishing, showcasing the increasing sophistication and personalization of phishing attacks \cite{Barracuda2022}. 

There have been a handful of works over the years that analyze large scale phishing email campaigns, with real-world phishing adversaries targeting a larger number and diversity of targets \cite{parsons2015design}. The work of Ho et al. \cite{ho2019detecting} focused primarily on the characterization of lateral phishing. Based on a massive dataset of 113 million employee-sent emails from 92 enterprise organizations, the study provided a detailed analysis of lateral phishing attacks. The researchers developed a classifier for detecting these attacks, and their findings shed light on the methods and behaviors of attackers. They discovered that most attackers rely on generic phishing content rather than crafting personalized attacks, indicating an opportunistic approach. Despite the lack of sophistication in content, these attacks were successful, with a significant percentage leading to the compromise of additional employee accounts. The work of Lain et al. \cite{lain2022phishing}, researchers conducted a 15-month phishing experiment with over 14,000 participants in a company. The study aimed to understand employee vulnerability to phishing, assess the effectiveness of phishing warnings and training, and explore the potential of crowd-sourced phishing detection. Key findings revealed that email warnings were effective in reducing phishing susceptibility, and that employees can be used as a collective phishing detection mechanism. As opposed to the previous two studies, the work of Oest et al. \cite{oest2020sunrise} focused on the detection and lifecycle of phishing email attacks. Their work proposed a framework that tracked the entire progression of phishing campaigns, from their online initiation to the compromise of accounts. Key findings include that the average phishing campaign lasts only 21 hours, with at least 7.42\% of visitors surrendering their credentials and subsequently experiencing fraudulent transactions. A notable aspect was that a small subset of campaigns accounted for the majority of victims.

Recently, concerns around LLMs being used for adversarial purposes, particularly in the context of phishing attacks, have risen significantly. A recent report by Google Cloud underscores that LLMs can be used to create highly legitimate-seeming content, including voice and video, making it more challenging to identify misspellings, grammar errors, and cultural inaccuracies in phishing emails and messages \cite{googlecloud2024}. There has also been an observed increase in the volume of phishing attackcs since the launch of ChatGPT, however the attack tactics and types of attacks have remained the same \cite{barracuda_networks_2024}. LLMs can also effectively translate and refine content, further complicating the detection of phishing attempts based on language use. The report points out that with generative AI, attackers can scale their operations and target a wide range of individuals with personalized and convincing emails, using data like names, organizations, job titles, and even health information. The work of Hazell et al. studied the effectiveness of LLMs toward creating realistic and cost-effective spear phishing emails to over 600 British Members of Parliament \cite{hazell2023large}. Their work also notes the possibility of bypassing safeguards in LLMs through basic prompt engineering, and emphasize the need for robust governance measures to prevent misuse. Another recent study \cite{sharma2023well} involved involved two groups of participants, exposed to either human-crafted or GPT-3 crafted phishing emails. Participants were asked to determine the authenticity of the emails, with feedback provided in the second round. The results indicated that human-crafted emails were more effective in deceiving people than those created by GPT-3, even after participants received training across various cognitive biases. Interestingly, participants felt more confident identifying phishing emails when they were human-crafted compared to those generated by GPT-3. Some recent work also explored how LLMs can be used to rewrite phishing emails to help bypass traditional email phishing filters \cite{afane2024next}. Beyond phishing emails, there has also been work that utilizes LLMs to generate the source code of phishing websites \cite{roy2024chatbots}. As a response to these growing concerns, there has been an increased focus on detecting LLM-generated content \cite{mitchell2023detectgpt, tang2024science, bethany2024deciphering}. Additionally, researchers are exploring the use of LLMs themselves to help detect phishing attempts \cite{li2024knowphish}.

While the existing large-scale phishing studies provide comprehensive insights into traditional phishing techniques and organizational vulnerabilities, they notably overlook the emerging threat posed by LLMs in phishing. Studies like those by Ho et al. \cite{ho2019detecting}, Lain et al. \cite{lain2022phishing}, and Oest et al. \cite{oest2020sunrise} offer valuable understanding of phishing campaign lifecycles, employee susceptibility, and detection mechanisms, yet they predate the rise of LLMs in cybersecurity threats. On the other hand, recent works focusing on LLM threats in phishing, such as those by Hazell et al. \cite{hazell2023large} and Sharma et al. \cite{sharma2023well}, while crucial in highlighting the potential misuse of LLMs for crafting sophisticated phishing emails, do not explore these threats on a large scale. These studies primarily concentrate on the theoretical aspects and potential implications of LLMs in phishing, hypothesizing about their effectiveness in creating realistic and tailored phishing content. However, they fall short in empirically validating the actual impact of LLM-crafted phishing across organizations and in understanding how individuals within these organizations behave in response to such advanced attacks. This gap underscores a critical need in phishing research: to integrate the nuanced understanding of LLM threats into broader, organizational-scale studies to comprehensively assess the evolving landscape of phishing attacks.

\section{Organization Security Incident Investigation: A Motivational Study}
\label{security_incident}

To contextualize our focus on studying LLMs for lateral phishing, we outline the cyber security threats the organization encounters based on official incident data provided by the organization's Cyber Operations Team, highlighting phishing as a predominant and financially damaging cyber threat. Subsequently, we detail a specific instance of lateral phishing that the organization experienced earlier in the year, documented in the team's security records. Building on this empirical evidence, we draw upon quantitative insights and threat intelligence from the organization's cyber operations team, which inform the design of our phishing experiments and shape the research questions we aim to answer through this study.

\subsection{Lateral Phishing and Impact}
\begin{figure}[]
    \centering
    \includegraphics[width=1\linewidth]{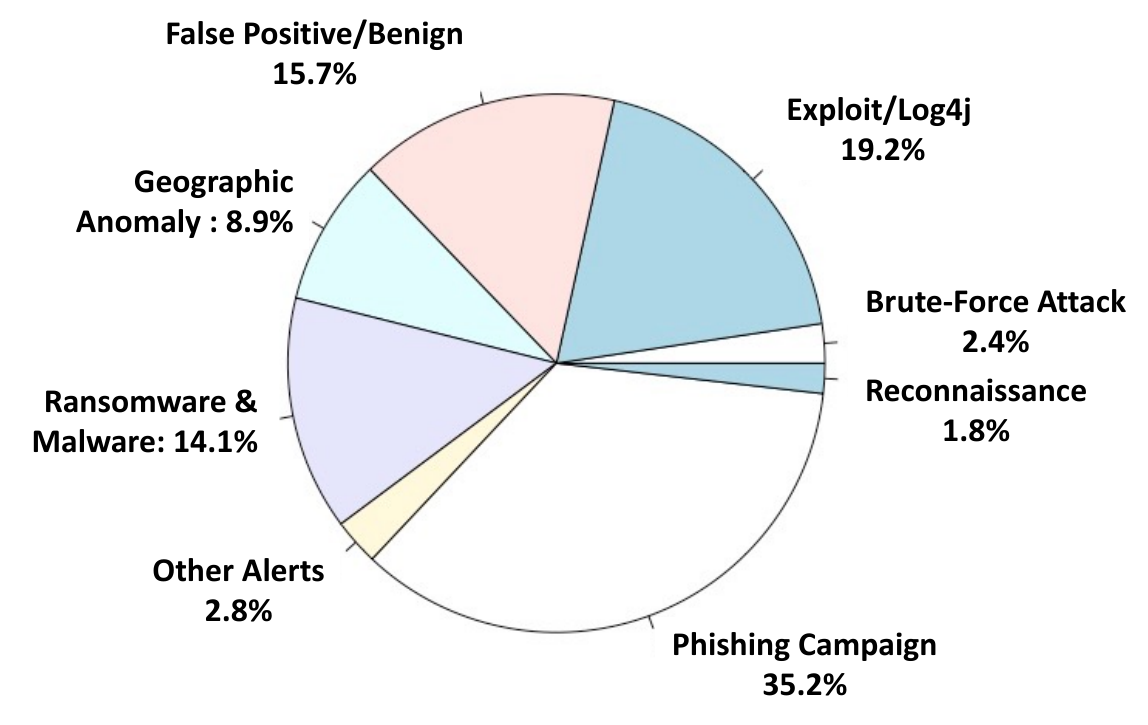}
    \caption{Breakdown of security incidents at the organization in 2023 with phishing campaigns constituting 35.2\% of total incidents.}
    \label{fig:security_incident}
\end{figure}

To motivate why we focus on lateral phishing email attacks, we first examine the existing state of information security within the organization and then conduct a comprehensive analysis of cyber security incidents from 2022-2023. Security incidents in this context, are part of daily operations that typically include tasks that are of low to moderate impact on systems and operations. These incidents require prompt and diligent attention. Although they do not immediately signal a breach, their significance escalates if not addressed efficiently. The cyber operations team takes a proactive approach to investigating these incidents, which often emerge as alerts for unusual login attempts, detected malware or viruses, correlations in event logs, and minor deviations from policy standards. Timely and effective management of these incidents is crucial in preventing their escalation into more serious situations. As depicted in Figure \ref{fig:security_incident}, the period witnessed roughly 13,000 security incidents. Notably, 35.2\% of these incidents were classified as phishing email campaigns, constituting the majority of all security incidents. Further analysis revealed a concerning trend in phishing incidents. In 2022, there were 1,947 incidents, which escalated to 2,245 in 2023, marking an approximate 15\% increase in just one year. Over the last 60 days, our security infrastructure automatically detected 18,693 phishing attempts. Despite this, 934 attacks eluded the automatic filters and led to successful phishing incidents. This data indicates that around 5\% of phishing attacks went undetected by our existing security measures and resulted in a phishing incident.

\begin{table}[b]
\centering
\resizebox{1\columnwidth}{!}{
\begin{tabular}{r|r}
\specialrule{1pt}{0pt}{-1pt}
\textbf{Category} & \textbf{Count} \\
\specialrule{1pt}{0pt}{0pt}
Blocked at the Edge & 2,247,769 \\
Total Email Processed & 38,855,303 \\
Forward Email Rules & 5,709,107 \\
Blocked Emails with Malware & 8,189 \\
Blocked Phishing Emails & 752,087 \\
Blocked Spam Emails & 2,136,707 \\
Blocked Malicious URLs & 9,342 \\
Retroactive Phishing Emails Removed & 40,944 \\
Retroactive Emails with Malware Removed & 251 \\
Retroactive Spam Emails Removed & 7,894 \\
Total Email Delivered & 30,180,782 \\
\specialrule{1pt}{0pt}{-1pt}
\end{tabular}
}
\captionsetup{skip=10pt}
\caption{Summary of last 90 days inbound Emails}
\label{tab:mail_summary_60days}
\end{table}


We also closely examine the last 90 days of Inbound Mail, as detailed in Table \ref{tab:mail_summary_60days}, a period that aligns with the duration of our phishing exercise. Throughout this time, the organization processed over 38 million emails, successfully delivering 30 million to user mailboxes. Close to 3 million emails were blocked, and 40,944 phishing emails were subsequently removed, highlighting the ongoing challenge of phishing attempts bypassing initial security protocols. The cyber operations team's investigation into each phishing campaign involves a detailed analysis, executed by a combination of personnel and automated systems. This typically amounts to more than 30 hours of collaborative work, drawing on resources from all levels of the IT organization. The cost associated with soft labor hours for addressing these phishing incidents is estimated at \$4.8 million over the past two years. Beyond the \$4.8 million figure, phishing poses additional, financial burdens that warrant recognition. Interruptions stemming from phishing attacks precipitate productivity losses, a substantial factor not encompassed within the direct cost calculation. Furthermore, there are indirect expenses linked to the consequences of these security breaches, including potential data loss, legal ramifications, and harm to the organization's reputation. These collateral costs, though not explicitly quantified in the \$4.8 million, place a strain on the organization's resources. Equally important is the financial commitment towards bolstering the security infrastructure as a response to the phishing threat. The investment in advanced security tools, sophisticated detection platforms, and continuous cyber awareness training forms a critical part of expenditures due to the threat of phishing. 



To enhance the relevance of our study on lateral phishing emails, we also underscore a recent phishing incident within our organization. In this lateral phishing incident, seven accounts were impacted, resulting in a combined effort of more than 2,000 soft labor hours and an estimated cost of \$73,500 to the organization. This attack specifically targeted high-ranking individuals, employing a sense of urgency in the email content and using cunning methods to propagate the attack, including creating inbox rules so the attacks would hide in plain sight. The primary aim was to harvest credentials, with a focus on diverting financial transactions to an overseas account. This sophisticated method underscored the urgent need to improve defenses against such attacks. It also provided a valuable opportunity to assess the effectiveness of our current training protocols. In response, the organization revised its monthly training modules and invested in an advanced phishing detection system that utilizes behavioral analytics to more efficiently detect various phishing schemes.

\subsection{Social Engineering Observations}

Social engineering tactics play a pivotal role in the effectiveness of phishing campaigns, as observed by our organization's cyber operations team. These insights reveal how attackers manipulate human psychology, often through seemingly legitimate offers or urgent requests, to trick individuals into compromising their security. One significant trend observed in lateral phishing attacks within the organization involves targeting recent contacts, especially direct reports, when the compromised account belongs to a supervising employee, validating observations on previous research on lateral phishing attacks \cite{alabdan2020phishing, ho2019detecting}. This method adds an extra layer to the attack, leveraging the existing trust within professional hierarchies. Another observation in existing phishing attacks faced by the organization are "timely" attacks that take advantage of publicly available knowledge to craft contextually relevant emails. Using the example of spring break, attackers might send emails promising exclusive vacation deals or last-minute bookings at significantly reduced rates to lure recipients into clicking on malicious links. Another notable example observed by the cyber operations team is phishing emails disguised as job opportunities, primarily targeting students. These emails cunningly use external links or attached Word documents as bait. In a broader context, phishing emails regarding Duo MFA, credential renewals, or resets are crafted with urgency to deceive recipients across the organization. Another widespread tactic involves falsely alarming users about the de-provisioning of their email accounts if they fail to take immediate action. Faculty and staff are often the targets of these sophisticated phishing attempts. These include spoofed login pages from recognized domains (such as r20.rs6.net) and emails with forged usernames from external domains, employing lateral phishing strategies. To enhance their deceptive appearance, many phishing emails include images linked to the organization or Microsoft, exploiting the trust associated with these entities. This integration of social engineering techniques into phishing campaigns underscores the need for continual vigilance and education in cyber security practices.



\subsection{Security Incident Conclusion}
\label{sec:Security_Incident_Conclusion}

The phishing email templates created for our comparative study are significantly influenced by the expert insights of our organization's cyber operations team. In this research, we compare human and LLM phishing capabilities on three scenarios grounded in the team's observations of past successful phishing campaigns. The first scenario represents a situation wherein a supervisor, leveraging the established rapport, emails their direct report, soliciting the execution of a specific task. This information is publicly available through organizational charts public directory and represents realistic information an attacker might use in lateral phishing. The second scenario unfolds through a more universally addressed email (e.g., events@ORG\_NAME.edu), urging the recipient to undertake a particular action in relation to an impending event. The final scenario presents an interaction from an employee's account, previously unassociated with the recipient, prompting an action aligned with a commonly observed phishing tactic within our organization (such as a fictitious job offer).


The key observations from the security incident team that we use to drive our investigation can be summarized as the following:
\begin{itemize}
    \item Lateral phishing emails often mimic communications from a supervisor to their direct reports, presenting a deceptive appearance of legitimacy.
    \item Many phishing attacks disguise themselves as urgent internal communications, impersonating various departments such as HR, Finance, Legal, IT, or Event Planning.
    \item Instances have been noted where mass messaging is executed through compromised internal email accounts.
    \item The crafting of phishing emails may involve the use of LLMs.
\end{itemize}

\subsection{Research Questions}
\label{sec:research_questions}

\noindent \textbf{RQ1: How does the effectiveness of lateral phishing emails generated by Large Language Models compare to that of those crafted by humans within an educational organization?} Understanding the effectiveness of lateral phishing attacks is critical to assessing organizational security vulnerabilities. This research question is novel as it investigates a large educational institution, contrasting with previous large-scale phishing research which predominantly focuses on commercial organizations. Our approach offers a unique perspective compared to the findings presented in previous prominent large scale phishing studies \cite{ho2019detecting, lain2022phishing}. Moreover, this question is crucial because it addresses the emerging threat of LLM-generated phishing emails, which can potentially lower the barrier for attackers by generating convincing, error-free content. By comparing the effectiveness of LLM and human-generated phishing emails, we can better understand the extent of this threat and inform more targeted defense strategies. Notably, our study compares LLM-generated emails against those crafted by a communications professional, representing the upper end of human ability in creating persuasive and coherent messages. This ensures a rigorous evaluation of the relative effectiveness of LLM-generated phishing emails against high-quality human-crafted emails.

\noindent \textbf{RQ2: Which job roles and department affiliations exhibit the highest risk of falling victim to phishing emails, as evidenced by elevated click-through rates, and what patterns can be discerned to identify these high-risk groups?} Identifying the job roles and department affiliations most susceptible to phishing attacks is essential for developing targeted awareness campaigns and allocating security resources effectively. By analyzing click-through rates across different job roles and departments, we can uncover patterns that may indicate specific vulnerabilities or knowledge gaps. This information is valuable for tailoring training programs and implementing more granular security measures to mitigate risks associated with high-risk groups. Furthermore, understanding these patterns can help organizations identify potential insider threats and develop strategies to address them proactively.

\noindent \textbf{RQ3: What are the underlying reasons for individuals to engage with lateral phishing emails?} Exploring the reasons and motivations behind individuals' engagement with phishing emails is crucial for developing effective user education and awareness programs. By understanding the factors that influence users' decision-making processes when faced with phishing attempts, organizations can design more targeted and persuasive training materials to help users identify and avoid these threats. Insights gathered from this investigation can inform the development of more comprehensive and adaptive defense strategies that account for the evolving nature of phishing attacks.

\section{Threat Model}
\label{data}

The proliferation of LLMs marks a significant shift in the domain of lateral phishing, with these models introducing complexities of an unprecedented scale. This paper considers two roles: (1) Adversarial agents who harness compromised email accounts alongside LLMs to forge highly individualized and persuasive phishing communication. These communications are carefully tailored to target specific individuals within an organization \cite{ayoobi2023looming}, with the ultimate goal of illicitly obtaining sensitive information and breaching secure systems \cite{thomas2017data}. The spectrum of adversaries ranges from solitary hackers to state-sponsored groups \cite{bossetta2018weaponization}. There is inherent trust that employees place in communication that appears familiar. This trust allows adversaries to craft and insert phishing messages into an organization’s communication flow undetected, exploiting this implicit trust to their advantage \cite{silic2016dark, hillman2023evaluating}. (2) Targets who receive these carefully engineered phishing communications. These targets span across all levels of hierarchy and profiles, and they may not always possess the necessary cybersecurity awareness to identify and avoid these sophisticated phishing attempts.
This study assumes that perpetrators have unauthorized access to both email accounts and LLMs. These tools enable them to create highly personalized and convincing phishing messages that utilize trust and create urgency, often resulting in the unintended release of sensitive information or actions that undermine the security of the targeted organization. LLMs can generate text that closely resembles genuine human writing. Furthermore, there is a common overestimation of the level of cybersecurity knowledge within an organization's staff, ignoring the reality that awareness and training in cybersecurity vary widely among employees.

\section{Lateral Phishing Simulation Infrastructure}
\label{infrastructure}
\begin{figure*}[t]
    \centering
    \includegraphics[width=0.8\linewidth]{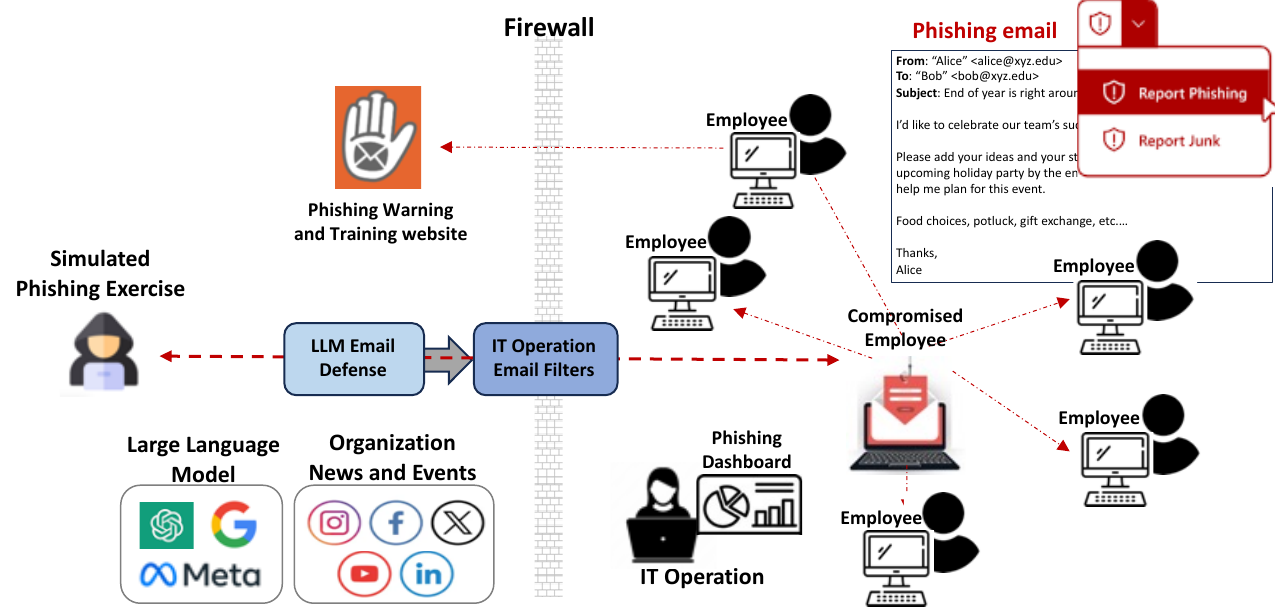}
    \caption{Schematic of Red-teaming infrastructure for lateral phishing simulations. This diagram illustrates the setup used for simulating phishing attacks within the organization to test the response efficacy of employees and IT systems.}
    \label{fig:organization}
\end{figure*}

\begin{figure}[b]
    \centering
    \includegraphics[width=1\linewidth]{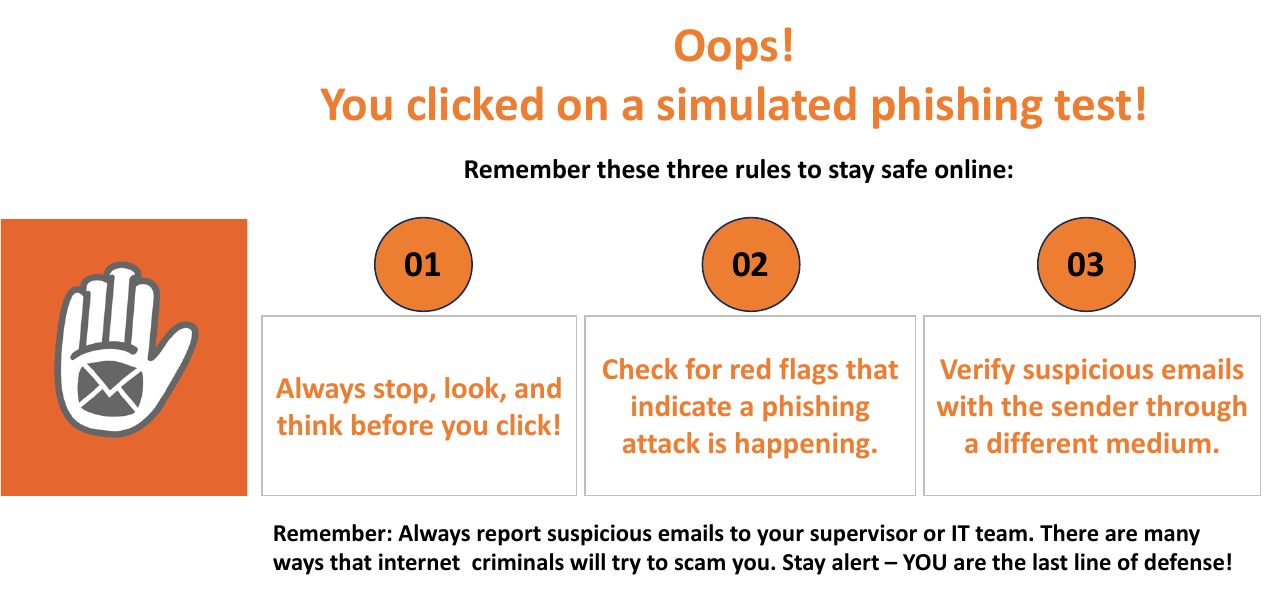}
    \caption{The informational warning message for our simulated phishing tests, detailing immediate steps to follow after clicking on a phishing link.}
    \label{fig:training}
\end{figure}

Figure~\ref{fig:organization} illustrates the setup of the red-teaming infrastructure used for lateral phishing simulations. The infrastructure consists of several key components working in concert. For the purposes of the red-teaming exercise, the simulated phishing emails bypass organizational email filters, mimicking either a compromised employee account or an organizational email address. The cyber operations team leverages their phishing simulation dashboard to incorporate organizational context such as employee names into the phishing emails. Additionally, in the context of this study, organization news from public sites were used to identify relevant upcoming events which would be the subject of one of the types phishing emails we studied. These simulated phishing emails are then distributed to targeted employee mailboxes. The cyber operations team records all employee interactions with the phishing emails, from whether the email was opened, to the link having been clicked, to having input login credentials into the phishing link. If an employee enters their login credentials through the phishing link, they are immediately redirected to a warning page (shown in Figure~\ref{fig:training}) that informs them they have participated in a simulated phishing test.

\subsection{Stakeholder Roles}

\subsubsection{Research Role}
For this study, we leveraged the already existing
phishing awareness campaign infrastructure to test our research questions. We partnered with the organization's cyber operations team and acted as advisors to administer the experiment. We worked with the university's cyber operations team prompt the LLM and human to develop effective phishing email templates. Researchers also acted as scientific advisors to guide the design of the experiments and research questions. Researchers only had access to anonymized data from the cyber operations team to perform analysis.

\subsubsection{IT Cyber Operations Role}
The cyber operations organization is strategically organized into three tiers: Identify and Protect, Threat Intelligence, and Respond and Recover. The team is committed to raising the organization's security posture and maintain a state of readiness within the institution. To ensure confidentiality, integrity and availability of institutional data and systems in the ever-evolving cyber-threat landscape, the organization emphasizes policy enforcement and compliance to safeguard intellectual property, proactively monitors and assesses cyber-threats keeping the institution alert, and mitigates risk of security incidents to ensure normal operations. 

Central to this cybersecurity framework is an overarching emphasis on training and awareness, especially against social engineering types of attacks. Monthly phishing simulation exercises cultivate an informed employee workforce and keep cybersecurity insurance premiums low. Weekly communication regarding scams and red flags raises the institutions defense mechanisms against sophisticated cyber-threats. It is a constant balancing act of threat detection and response activities, systems patching and risk mitigation strategies complemented by phishing simulations and training.

\subsubsection{Study Participants}
In this study, 9129 employees were subjected to the phishing exercise. Among these employees, 1386 held supervisory roles. This distinction is important because one of the email templates used in the study relied on lateral phishing attacks, where the email was addressed from a supervisor to a direct report, exploiting the trust and familiarity inherent in this common type of organizational relationship. We also analyzed the results of the phishing experiment by department and role. The organization had 302 unique departments, which were manually mapped to 10 department categorizations. The study encompassed a range of employee types across the university. There were 1,219 unique job roles, which were also manually mapped to 9 job role categorizations.

\subsection{Phishing Process}

The organization leverages a comprehensive phishing training platform designed for realistic cybersecurity training exercises. Phishing campaigns are normally conducted over one to three days and emails are sent during working hours. Employees are sampled in random order and templates are designed to match specific criteria to test the employee's ability to identify sophisticated phishing attacks. The help desk is particularly impacted during these exercises as they receive calls and reports of suspicious emails. The help desk is not notified of the time frame phishing exercises are conducted to reflect real-world scenarios. This also presents the opportunity to refine their processes and procedures to handle unexpected surge of tickets. The organization also leverages phishing detection tools in several other platforms that cover signature based and behavioral based detection and alerts as well as the ability to orchestrate a full investigation of backtracking the origin in emails of interest. The organization also conducts annual mandatory security training covering a broad range of security topics to ensure employees stay abreast of latest security threats and tactics. To further assist the cyber operations team in reducing the time to respond and resolve phishing incidents, the organization leverages a threat intelligence solution that provides extensive visibility into various entry points to reveal high-impact events and attack insights to further strengthen the team's response mechanisms.

\subsection{Phishing Emails}

To answer our research questions, we devised a standardized methodology for formulating the email templates. Both the human participant, a professional with a background in communications representing the upper-end of human ability in crafting persuasive and coherent messages, and the LLM, specifically ChatGPT GPT-4, were given identical prompts to generate phishing emails. To maintain consistency and fairness in the evaluation, the initial drafts produced by both the human and the LLM were subjected to a review process by the research team. This process involved making minimal adjustments to the human-composed email to align it closely with its LLM counterpart in terms of the sender's identity, the email's subject matter, the degree of personalization, and the conveyed sense of urgency. Utmost care was taken to ensure that the writing style of each email was maintained, preserving the unique characteristics of human and LLM-generated content. This approach was carefully chosen to ensure that any observed variances in the effectiveness of the phishing attempts could be attributed primarily to the intrinsic differences between human and LLM-generated content, rather than extraneous factors. Across all emails, a sense of urgency was required as part of the prompt to generate the emails, as this is a common tactic employed in phishing attempts to pressure recipients into taking immediate action without thoroughly considering the consequences. We tested three types of lateral phishing: emails from a supervisor, organizational-level emails, and emails from another internal employee who the recipient does not know, to assess the effectiveness of different attack vectors and identify potential vulnerabilities within the institution's hierarchy and communication channels. For all phishing email templates generated, the recipient is asked to complete some action by clicking a link in the email so that we could record the recipient's vulnerability to clicking a phishing link and entering data or credentials in the web interface the link leads to. In the following subsubsections, we outline the prompts given to the LLM and human to generate the lateral phishing email templates and reasoning behind why these topics were chosen.

\subsubsection{Supervisor to direct report}

\noindent\textbf{Prompt}: "Write a generic email from a supervisor to direct report, where the direct report is instructed to click some link with some urgency to enter some information about professional development."

In this scenario, we asked both the LLM and the human to compose an email from a supervisor to their direct report, using the prompt above. The email instructs the recipient to click on a link with urgency to enter information about professional development. This topic was chosen because it represents a plausible situation in which a supervisor might make such a request of their direct report. The inherent power dynamics and trust within the supervisor-direct report relationship could potentially make this type of phishing attempt highly effective, as employees may feel compelled to comply with their supervisor's instructions without questioning the legitimacy of the request. In subsequent sections we describe this email topic as "Supervisor to Direct Report".

\subsubsection{Organizational-level email on a timely event}

\noindent\textbf{Prompt}: "Write a generic email on some timely internal communications where the sender would be from some organizational level email address, where the email asks the recipient to enter some information in a link with some urgency. The topic of the email should about making reservations for eyewear relating to the upcoming solar eclipse."

For this topic, the LLM and human were tasked with writing a generic email using the prompt above. The email is sent from an organizational-level email address, asking the recipient to enter information via a link with urgency. The specific subject of the email was making reservations for eyewear related to an upcoming solar eclipse. This topic was selected because the solar eclipse was a well-known event that was going to occur, and the institution was known to be hosting events related to the eclipse. Moreover, the organization had previously experienced phishing attacks that capitalized on timely events, making this a relevant and realistic scenario to test the effectiveness of LLM and human-generated phishing emails in exploiting current events and institutional context. In subsequent sections we describe this email topic as "Timely Phishing".

\subsubsection{Common Phishing Email from Internal Employee Account}

\noindent\textbf{Prompt}: "Write a generic email about extending a job offer where the recipient would be doing some remote work and is asked to click on a link to enter some information with some urgency."

In this case, both the LLM and human were asked to compose a generic email using the prompt above. The email is about extending a job offer, where the recipient would be doing some remote work and is asked to click on a link to enter information with urgency. According to the organization's cyber operations team, job offer phishing schemes are the most common type of phishing attack faced by the institution. By using this topic, we can assess the effectiveness of LLM and human-generated phishing emails in replicating a prevalent real-world threat. Additionally, the scenario involves an email from an unfamiliar internal employee, testing the recipient's ability to discern the legitimacy of the request without the inherent trust that may be present in established relationships within the organization. In subsequent sections we describe this email topic as "Common Phishing".


\subsection{Phishing Training and Warning}
The annual security awareness training that employees are automatically enrolled and must complete reinforces the institution's policy which has awareness of threats embedded with the acceptable use of computing resources. The department also sends weekly "Scam of the Week" emails with advisories on current trends and brand these use cases with "Stop, Look, and Think. Don't be fooled." at the end of each advisory security safety tips are outlined to watch for emails that contain a sense of urgency, any prompts for a call back, or suspicious context and grammar structure. Employees who are identified to enhance their cybersecurity awareness based on the monthly phishing exercise results are given a training opportunity and two weeks to complete it. Timely training has been shown to raise awareness to these phishing threats \cite{reinheimer2020investigation}. They get a periodic reminder and if the training is incomplete then an email goes to their manager. Upon completion of their assigned training they receive a thank you email for completing it and for doing their part to keep the organization safe from cyber attacks. When designing these phishing exercises the following criteria are evaluated: Prior actual security incident types, current trends, Red Flag indicators such us external, internal, sense of urgency, subject, attachments with macros, spoofed domains or email addresses, fake login portals and landing domains.

\begin{table*}[]
\centering
\resizebox{\textwidth}{!}{
\begin{tabular}{r|r|r|r|r|r}
\specialrule{1pt}{0pt}{-1pt}
\textbf{Email Author} & \textbf{Email Topic} & \textbf{Email Recipients} & \textbf{Emails Opened} & \textbf{Link Clicked} & \textbf{Data Entered} \\
\specialrule{1pt}{0pt}{0pt}
Human                  & Supervisor to Direct Report & 1541 & 955 (61.97\%) & 334 (21.67\%) & 164 (10.64\%) \\
LLM                    & Supervisor to Direct Report & 1490 & 919 (61.68\%) & 318 (21.34\%) & 165 (11.07\%) \\
Human                  & Timely Phishing & 1591 & 574 (36.08\%) & 106 (6.66\%) & 40 (2.51\%) \\
LLM                    & Timely Phishing & 1479 & 597 (40.37\%) & 153 (10.34\%) & 62 (4.19\%) \\
Human                  & Common Phishing & 1520 & 831 (54.67\%) & 68 (4.47\%) & 18 (1.18\%) \\
LLM                    & Common Phishing & 1508 & 715 (47.41\%) & 73 (4.84\%) & 14 (0.93\%) \\
\specialrule{1pt}{0pt}{0pt}
                       & Total Human Written & 4652 & 2360 (50.73\%) & 508 (10.92\%) & 222 (4.77\%) \\
                       & Total LLM Generated & 4477 & 2231 (49.83\%) & 544 (12.15\%) & 241 (5.38\%) \\
\specialrule{1pt}{0pt}{0pt}
                      &  Total                    & 9129 & 4591 (50.29\%) & 1052 (11.52\%) & 463 (5.07\%) \\
\specialrule{1pt}{0pt}{-1pt}
\end{tabular}
}
\caption{Comparison of human-crafted and LLM-generated phishing email effectiveness. Percentages are relative to the number of email recipients.}
\label{tab:email_performance}
\end{table*}

\subsection{Measurement}

In conducting phishing exercises, we closely monitor employee responses, gathering essential metrics such as Sent, Opened, Clicked, Replied, Attachment, Data Entered (without saving credentials), and Reported. By aggregating these metrics with data from previous exercises, we formulate a risk score to evaluate the organization's security posture and assess the susceptibility of our employee base to phishing attacks. Employees who enter data in these exercises are promptly enrolled in security awareness training, with a completion deadline of two weeks. Additionally, this risk score is benchmarked against industry data to provide insights, track historical trends, and gauge the maturity level of the organization's cybersecurity posture.

In measuring the effectiveness of these attacks, our focus centers on two pivotal metrics: click-through rates and data exfiltration success. Click-through rates are determined by the percentage of recipients who click on links or attachments in the phishing emails, serving as an indicator of the email's ability to persuade the recipient to take action \cite{steves2020categorizing}. The data exfiltration success, which in our context refers to the rate at which recipients enter data, provides insight into the effectiveness of LLM-generated emails in compelling recipients to divulge sensitive information. Data exfiltration, particularly in the context of recipients entering data, represents one of the most critical aspects of a phishing attack, as it often involves the surrender of sensitive information, such as login credentials. This breach is particularly serious because once attackers gain access to login details, they can infiltrate the organization's network, access confidential information, and potentially cause additional harm to other users. Therefore, the rate at which recipients are tricked into entering such sensitive data is a key indicator of the severity and potential impact of a phishing attack \cite{sabir2021machine, ullah2018data, tejaswi2022leaky}.

\subsection{Ethics, Safety and Privacy}
The primary goal of the phishing simulation exercises is to enhance the employees' ability to identify and mitigate potential Cyber threats effectively, conduct those in a supportive environment that aids in better preparedness against cyber attacks, and to emphasize learning and skill development in the workforce. In designing a phishing simulation test, the Office of Information Security evaluates collective threat intelligence information and discusses the goals of the exercise with the cyber operations team. Several factors are considered: raising awareness, improve the overall security posture of the organization, the impact to the individual, the timing and frequency of the simulation emails, phishing template design, compliance to existing policies, privacy laws and regulations (CCPA, FERPA, PII, TGC.s2054.519(b), TCC, TSPA) and the level of disruption in normal operations to the organization. This is particularly challenging when conducting Man-in-the-Middle (MitM) phishing simulation exercises, as the population is large and diverse. The team considers the ethical implications in spoofing domains or individual's accounts and contrasts those with real-world phishing campaigns. Separation of duties is a critical centerpiece of the process to ensure integrity of the data and the process itself, and to minimize the risk of mishandling sensitive information. During an exercise, any data that a participant enters to a login form (credentials harvesting) or interception simulation is not stored. Only the aggregated results of the data collected and Risk Score calculated during a monthly test are presented to executive leadership to ensure individuals' privacy is protected. Phishing simulation exercises can raise employee concerns, so it is important to have an empathetic and informative approach, and present those as valuable learning opportunities.

In this study, our team collaborated with the University's cyber operations team, which had previously been conducting exercises using phishing emails to enhance awareness and educate users about the risks associated with such emails. The cyber operations team was responsible for managing all personally identifiable information (PII), ensuring its anonymization before providing the researchers with statistical data and other qualitative insights. Upon reviewing our research protocols, the IRB of our institution determined that our project did not fall under the regulatory criteria of the DHHS or the FFDA. Consequently, it did not require further oversight from the IRB. While there could be concerns about the lack of informed consent, these phishing simulation exercises are currently already conducted throughout the organization on a regular basis as a part of standard organizational security practices, with researchers simply helping to inform the methodology to generate emails. The cyber operations team communicated the details of the proposed lateral phishing exercise with IT governance in the University system to obtain approval to conduct the exercise, where the results of the exercise could help inform the organization quantify the potential threat of LLMs towards phishing within the organization and potentially inform security practices and training in the future. To reduce the psychological impact, the cyber operations team established a trusted agents committee that meets prior to the launch of phishing campaigns. This committee convenes before phishing campaigns and engages in a discussion forum where members review and approve the email templates, offer feedback, and adjust accordingly. The committee consists of more than eight members representing different areas including chief of staff, risk management and legal affairs, College of Business, Academic affairs operations, Human Resources, strategic enrollment and student affairs, and research integrity, in order to limit potential harms regarding the phishing campaign.

\section{Experiments and Discussion}
\label{discussion}

\subsection{RQ1: Human vs LLM Lateral Phishing emails}
\label{sec:human_vs_llm}

\begin{table*}[]
\centering
\resizebox{\textwidth}{!}{
\begin{tabular}{r|r|r|r|r|r}
\specialrule{1pt}{0pt}{-1pt}
\textbf{Email Author} & \textbf{Email Topic} & \textbf{Email Recipients} & \textbf{Emails Opened} & \textbf{Link Clicked} & \textbf{Data Entered} \\
\specialrule{1pt}{0pt}{0pt}
Human                  & Supervisor to Direct Report & 249 & 178 (71.49\%) & 47 (18.88\%) & 12 (4.82\%) \\
LLM                    & Supervisor to Direct Report & 241 & 176 (73.03\%) & 46 (19.09\%) & 20 (8.30\%) \\
Human                  & Timely Phishing & 225 & 120 (53.33\%) & 22 (9.78\%) & 12 (5.33\%) \\
LLM                    & Timely Phishing & 227 & 150 (66.08\%) & 40 (17.62\%) & 15 (6.61\%) \\
Human                  & Common Phishing & 228 & 159 (69.74\%) & 2 (0.88\%) & 1 (0.44\%) \\
LLM                    & Common Phishing & 216 & 126 (58.33\%) & 1 (0.46\%) & 0 (0.00\%) \\
\specialrule{1pt}{0pt}{0pt}
                       & Total Human Written & 702 & 457 (65.10\%) & 71 (10.11\%) & 25 (3.56\%) \\
                       & Total LLM Generated & 684 & 452 (66.08\%) & 87 (12.72\%) & 35 (5.12\%) \\
\specialrule{1pt}{0pt}{0pt}
                      &  Total                    & 1386 & 909 (65.58\%) & 158 (11.40\%) & 60 (4.33\%) \\
\specialrule{1pt}{0pt}{-1pt}
\end{tabular}
}
\caption{Comparison of human-crafted and LLM-generated phishing email effectiveness. Percentages are relative to the number of email recipients. Filtered down to only those who supervise at least one other employee the organization.}
\label{tab:supervisor}
\end{table*}

\begin{table}[b]
\centering
\resizebox{\columnwidth}{!}{
\begin{tabular}{r|r|r|r}
\specialrule{1pt}{0pt}{-1pt}
  & \textbf{open} & \textbf{click} & \textbf{enter} \\
\specialrule{1pt}{0pt}{0pt}
supervisor to direct report & 0.167071 & 0.222355 & -0.38169 \\
timely phishing & -2.44353* & -3.66803* & -2.59189* \\
common phishing & 3.994332* & -0.47943 & 0.688341 \\
\specialrule{1pt}{0pt}{-1pt}
all & 0.85823 & -1.8414* & -1.3299 \\
\specialrule{1pt}{0pt}{-1pt}
\end{tabular}
}
\caption{t-tests comparing human vs LLM emails across lateral phishing topics. Positive values indicate greater effect by human-authored emails; negative values by LLM-authored emails. Significant differences (*) are at p < 0.05 after False Discovery Rate adjustment.}
\label{tab:t-test_emails}
\end{table}

To answer Research Question 1, which compares the effectiveness of lateral phishing emails generated by LLMs and those crafted by humans within an educational organization, we analyze the data presented in Table \ref{tab:email_performance} and the t-test results for statistical significance of this comparison in Table \ref{tab:t-test_emails}.

\begin{table*}[t!]
\centering
\resizebox{0.9\textwidth}{!}{
\begin{tabular}{r|r|r|r|r}
\specialrule{1pt}{0pt}{-1pt}
\textbf{Department}                                         & \textbf{Email Recipients} & \textbf{Emails Opened}        & \textbf{Link Clicked}         & \textbf{Data Entered}         \\
\specialrule{1pt}{0pt}{0pt}
College of Business                                & 603             & 388 (64.34\%)        & 102 (16.92\%)        & 37 (6.14\%)          \\
College of Education and Human Development         & 480             & 252 (52.50\%)        & 45 (9.38\%)          & 23 (4.79\%)          \\
College of Engineering                             & 638             & 340 (53.29\%)        & 81 (12.70\%)         & 29 (4.55\%)          \\
College of Liberal and Fine Arts                   & 743             & 429 (57.74\%)        & 87 (11.71\%)         & 46 (6.19\%)          \\
College of Sciences                                & 1270            & 768 (60.47\%)        & 155 (12.20\%)        & 57 (4.49\%)          \\
College of Architecture, Construction and Planning & 166             & 88 (53.01\%)         & 22 (13.25\%)         & 12 (7.23\%)          \\
Honors College                                     & 509             & 288 (56.58\%)        & 74 (14.54\%)         & 36 (7.07\%)          \\
University College                                 & 315             & 168 (53.33\%)        & 40 (12.70\%)         & 15 (4.76\%)          \\
Administration                                     & 2685            & 1214 (45.21\%)       & 300 (11.17\%)        & 139 (5.18\%)         \\
Other                                              & 1720            & 656 (38.14\%)        & 146 (8.49\%)         & 69 (4.01\%)          \\
\specialrule{1pt}{0pt}{0pt}
Total                                              & 9129            & 4591 (50.29\%)       & 1052 (11.52\%)       & 463 (5.07\%)         \\
\specialrule{1pt}{0pt}{-1pt}
\end{tabular}
}
\caption{Effectiveness of phishing emails across different departments. Percentages are relative to the number of email recipients}
\label{tab:department_email_performance}
\end{table*}

Table \ref{tab:email_performance} provides an overview of the phishing email performance metrics for both human-crafted and LLM-generated emails across three different lateral phishing email topics: supervisor to direct report, timely phishing, and common phishing. The data shows that the overall percentage of emails opened, links clicked, and data entered are relatively similar between human-written and LLM-generated emails. For the "supervisor to direct report" scenario, the human-crafted email had a slightly higher percentage of emails opened (61.97\% vs. 61.68\%) and links clicked (21.67\% vs. 21.34\%), while the LLM-generated email had a slightly higher percentage of data entered (11.07\% vs. 10.64\%). In the "timely phishing" scenario, the LLM-generated email outperformed the human-crafted email across all metrics, with a higher percentage of emails opened (40.37\% vs. 36.08\%), links clicked (10.34\% vs. 6.66\%), and data entered (4.19\% vs. 2.51\%). For the "common phishing" scenario, the human-crafted email had a higher percentage of emails opened (54.67\% vs. 47.41\%), while the LLM-generated email had a slightly higher percentage of links clicked (4.84\% vs. 4.47\%) and a lower percentage of data entered (0.93\% vs. 1.18\%).

To determine the statistical significance of these differences, we conducted t-tests comparing the human-crafted and LLM-generated emails for each phishing email topic, shown in Table \ref{tab:t-test_emails}. Because performing multiple comparisons increases the risk of false positives, we applied the Benjamini-Hochberg False Discovery Rate (FDR) correction. This correction adjusts the p-values to control the proportion of type I errors (i.e., falsely identifying a difference as significant). Consequently, only comparisons with adjusted p-values less than 0.05 are considered statistically significant, meaning there is less than a 5\% probability that these observed differences occurred by chance. Specifically, for the 'timely phishing' scenario, the LLM-generated email performed significantly better than the human-crafted email in terms of emails opened (t = -2.44, p $<$ 0.05), links clicked (t = -3.67, p $<$ 0.05), and data entered (t = -2.59, p $<$ 0.05). In the 'common phishing' scenario, the human-crafted email had a significantly higher percentage of emails opened (t = 3.99, p $<$ 0.05), while no significant differences were observed for links clicked or data entered. For the 'supervisor to direct report' scenario, no statistically significant differences were found. These t-values indicate both the direction and magnitude of the differences, where larger absolute values correspond to stronger effects. The p $<$ 0.05 threshold confirms that the chance of these findings occurring randomly is below 5\%.

These findings suggest that LLM-generated phishing emails can be as effective as, and in some cases even more effective than, human-crafted phishing emails. The comparable performance of LLM-generated emails in the "supervisor to direct report" scenario highlights the potential threat of LLM phishing attacks, as they can exploit the trust and power dynamics within an organization. In conclusion, our analysis reveals that LLM-generated lateral phishing emails can be as effective as those crafted by humans within an educational organization. This finding underscores the need for organizations to adapt their cybersecurity strategies to address the growing threat of AI-powered phishing attacks and to invest in employee training and awareness programs that specifically target the unique characteristics of LLM-generated phishing emails.


Given the effectiveness of the lateral phishing emails from the supervisor to direct report, we further investigated the susceptibility of those who hold supervisory roles themselves since a single compromised account where they hold a supervisory role could lead to an effective propagation of an attack. Comparing the results of the phishing experiment for supervisors (Table \ref{tab:supervisor}) to the entire organization (Table \ref{tab:email_performance}), we observe that supervisors have significantly higher email open rates across all email topics. The most notable difference is in the "Timely Phishing" category, where supervisors have an open rate of 59.71\% compared to 38.23\% for the entire organization.
However, when examining link click rates and data entry rates, supervisors show varying levels of susceptibility depending on the email topic. For the "Supervisor to Direct Report" category, supervisors have lower link click (18.98\%) and data entry (6.56\%) rates compared to the entire organization (21.51\% and 10.86\%, respectively). This suggests that supervisors may be more cautious when receiving emails appearing to be from their own supervisors.
In the "Timely Phishing" category, supervisors have higher link click (13.72\%) and data entry (5.97\%) rates compared to the entire organization (8.44\% and 3.32\%, respectively). This indicates that supervisors may be more vulnerable to timely and relevant phishing attempts. 


\subsection{RQ2: Breakdown by job role and department affiliation}
\label{sec:demographics}

\begin{table*}[t!]
\centering
\resizebox{0.8\textwidth}{!}{
\begin{tabular}{r|r|r|r|r}
\specialrule{1pt}{0pt}{-1pt}
\textbf{Role} & \textbf{Email Recipients} & \textbf{Emails Opened} & \textbf{Link Clicked} & \textbf{Data Entered} \\
\specialrule{1pt}{0pt}{0pt}
Administration \& Leadership & 408 & 253 (62.01\%) & 45 (11.03\%) & 21 (5.15\%) \\
Athletics & 122 & 73 (59.84\%) & 12 (9.84\%) & 4 (3.28\%) \\
Faculty \& Research & 1875 & 1272 (67.84\%) & 217 (11.57\%) & 88 (4.69\%) \\
Finance & 202 & 72 (35.64\%) & 26 (12.87\%) & 11 (5.45\%) \\
Human Resources \& Payroll & 42 & 27 (64.29\%) & 6 (14.29\%) & 1 (2.38\%) \\
IT \& Information Security & 201 & 107 (53.23\%) & 27 (13.43\%) & 10 (4.98\%) \\
Student Roles & 3193 & 1604 (50.23\%) & 449 (14.06\%) & 216 (6.76\%) \\
Student Services \& Advising & 250 & 129 (51.60\%) & 28 (11.20\%) & 14 (5.60\%) \\
Other Roles & 2836 & 1054 (37.17\%) & 242 (8.53\%) & 98 (3.46\%) \\
\specialrule{1pt}{0pt}{0pt}
Total & 9129 & 4591 (50.29\%) & 1052 (11.52\%) & 463 (5.07\%) \\
\specialrule{1pt}{0pt}{-1pt}
\end{tabular}
}
\caption{Effectiveness of phishing emails across different organizational roles. Percentages are relative to the number of email recipients.}
\label{tab:role_interaction}
\end{table*}

To address RQ2, we analyzed the click-through rates across different job roles and department affiliations within the university. Table \ref{tab:department_email_performance} presents the combined data for each department's email activities, including the number of emails delivered, opened, links clicked, and data entered. The percentages provided are relative to the number of emails delivered to each department.
Comparing the percentages across departments, we observe that the College of Business had the highest percentage of emails opened at 64.34\%, followed by the College of Sciences at 60.47\%. The College of Architecture, Construction and Planning had the highest percentage of links clicked at 13.25\%, while the "Other" category had the lowest at 8.49\%. The Honors College and the College of Architecture, Construction and Planning had the highest percentages of data entered at 7.07\% and 7.23\%, respectively.
Table \ref{tab:role_interaction} shows the effectiveness of the phishing emails across different organizational roles. The percentages are relative to the number of email recipients for each role. Faculty \& Research had the highest percentage of emails opened at 67.84\%, while the "Other Roles" category had the lowest at 37.17\%. Human Resources \& Payroll had the highest percentage of links clicked at 14.29\%, followed closely by Student Roles at 14.06\%. Student Roles, which include positions typically filled by students such as graduate assistants, teaching assistants, research assistants, and student workers, also had the highest percentage of data entered at 6.76\%.
The College of Business and the College of Sciences was observed to have higher email open rates, while the College of Architecture, Construction and Planning and the Honors College had higher percentages of links clicked and data entered. Similarly, Faculty \& Research and Human Resources \& Payroll roles exhibited higher engagement with phishing emails compared to other roles. The high engagement of Student Roles with phishing emails is particularly concerning, as these positions are often filled by less experienced individuals who may be more vulnerable to social engineering tactics.

\subsection{RQ3: Motivation for engaging with phishing emails}
\label{sec:questionnaire}

To understand the underlying reasons and motivations for individuals to engage with LLM and human-written phishing emails, we conducted a questionnaire on 34 employees who entered data in the web interface after clicking the phishing link. This optional questionnaire was sent out to all 463 employees those who entered data, but only 34 completed the questionnaire. The questionnaire consisted of Likert scale questions and open-ended qualitative questions.

Likert Questions (Rating from 1-5: 1. Strongly Disagree, 2. Somewhat Disagree, 3. Neither Agree nor Disagree, 4. Somewhat Agree, 5. Strongly Agree):

\begin{enumerate}
\item "The fact that the email came from an internal (@ORG\_NAME.edu) email address influenced my decision to trust the email." (Average likert score: 4.77)
\item "I decided to interact with the email because I felt confident about the sender's identity." (Average likert score: 4.51)
\item "My perception of the email's relevance to my personal or professional interests significantly guided my decision to interact with it." (Average likert score: 4.35)
\item "My perception of the email's urgency (such as a warning, a deadline, or a request) significantly guided my decision to interact with it." (Average likert score: 3.43)
\end{enumerate}

The high average likert scores for questions 1 and 2 suggest that the sender's identity, particularly the fact that the email came from an internal email address, played a significant role in the respondents' decision to trust and interact with the email. The high score for question 3 indicates that the perceived relevance of the email to the respondents' personal or professional interests also influenced their decision to engage with the email. The lower score for question 4 suggests that while the perceived urgency of the email did influence the respondents' decision, it was not as strong a factor as the sender's identity and the email's relevance.

Open-ended Qualitative Questions:

\begin{enumerate}

\item "Which elements, if any, did you evaluate or scrutinize to assess the email's legitimacy?"

\medskip

Only one respondent reported not scrutinizing the email, despite the fact that this survey was only taken by those who entered data into the phishing link. This could indicate either a lack of forthcomingness due to shame or other social stigma, or that the respondents proceeded to enter data despite scrutinizing the email's legitimacy.

\medskip

Common elements scrutinized included the sender's email address and the wording of the email. Some respondents reported looking for errors (typos, misspellings, or strange alphanumeric symbols) in the email body, as these are common in many phishing attacks.

\item "What motivated you to engage with the email?"

\medskip

Respondents reported trusting the email since it came from an internal address and the link pointed to a domain name affiliated with the institution. This suggests an awareness of phishing but not of lateral phishing involving compromised internal accounts.

\medskip

The topic of the email was relevant to the recipient. For example, some respondents reported having discussed professional development with their supervisor earlier that day and then receiving the phishing template about professional development. Others found the timely email involving the solar eclipse convincing since they were aware of the institution's planned activities related to the event.

\medskip

Some respondents reported being busy and stressed, and the urgency of the email prompted them to follow the instructions without further consideration.

\medskip

Some respondents felt pressure to complete the task in a timely manner since the email appeared to come from their supervisor.

\medskip

\item "Is there anything else you'd like to share about your experience with the email for us to better understand your interaction with the email?"

\medskip

Respondents mainly shared their thoughts and suggestions about the phishing training itself. Some appreciated the training and believed it would increase their phishing awareness in the future, while others expressed displeasure and considered it useless or a waste of time.

\end{enumerate}

These findings provide valuable insights into the underlying reasons and motivations for individuals to engage with the lateral phishing emails. The respondents' trust in internal email addresses, the perceived relevance of the email content, and the sense of urgency conveyed in the emails were key factors influencing their decision to interact with the phishing emails. The results also highlight the need for more comprehensive phishing awareness training that addresses lateral phishing and the exploitation of compromised internal accounts. Although some respondents were vigilant in identifying characteristics of typical phishing emails, they seemed vulnerable to the well-crafted emails that employed internal email addresses, suggesting a gap in their understanding of these more sophisticated phishing techniques that could be exacerbated in the age of LLM powered phishing. Additionally, the findings underscore the importance of fostering a culture of cybersecurity awareness and providing employees with the necessary tools and knowledge to identify and report suspicious emails, even when under pressure or stress.

\subsection{Detection of LLM written phishing emails}

Tagging LLM generated emails may be a solution to be explored in future work to help mitigate the threat of LLM phishing emails. When an LLM-generated email is detected, proactive measures may be taken to mitigate its potential impact. One potentially effective strategy is to automatically tag such emails, informing the recipient of the potential risks. This tag would act as an immediate visual cue, alerting the recipient that the email has been flagged as being potentially LLM-generated, thus warranting extra scrutiny. Email tagging has been investigated in previous research as it augments users ability to more quickly categorize content \cite{nelson2011mail2tag, koren2011automatically}. Currently within the organization, a similar tagging methodology is used to indicate emails that come from outside the organization, assigning an [EXTERNAL] tag to all such emails. Such tags would not only serve as a direct warning but also reinforce the training employees receive, bridging the gap between theoretical knowledge and practical application. This dual approach of enhancing both technological defenses and employee vigilance could be part of a defense strategy against the evolving landscape of LLM-generated phishing threats. Future research could investigate email tagging for LLM-generated content to see if these tags could heighten user awareness to LLM phishing threats.

In order to investigate if LLM-generated phishing emails can be detected, we performed a small preliminary experiment using a recent LLM detection system that utilizes T5 \cite{flan-t5-xl} encoder embeddings in a Multilayer Perceptron (MLP) classifier \cite{bethany2024deciphering}. We built our dataset by taking 5,000 human-written email samples \cite{chakraborty_2023} and generating corresponding machine-written variants using the Vicuna 13B v1.5 model, thus creating paired examples to capture the subtle differences between human and LLM-generated texts. The detection system works by feeding each email into a frozen T5 encoder that extracts dense, semantically rich embeddings, which are then processed by an MLP that learns to classify the texts; our preliminary results show that, after fine-tuning on this domain-specific dataset, the  classifier achieves 98 F1 score on this test. It is important to note that this experiment is only a brief test within our broader study, and a more detailed investigation into advanced detection strategies remains a promising direction for future work.

While email tagging presents a promising defensive approach, several technical challenges must be addressed for effective implementation. First, detection mechanisms must balance sensitivity and specificity, where too aggressive detection could flag legitimate communications, and lenient systems might miss sophisticated LLM-generated phishing attempts. A fundamental technical challenge in the detection of LLM generated content lies in the detection of content from unknown or novel LLMs. While detecting generations from a specific, known LLM is relatively straightforward \cite{mitchell2023detectgpt}, real-world scenarios involve many different models (with tens of thousands of LLMs available on platforms like Hugging Face alone), making comprehensive coverage extremely difficult. This "in-the-wild" detection problem (identifying LLM-generated content without prior knowledge of the specific generating model) remains largely unsolved and represents a significant barrier for detecting LLM generated text. False positives could erode user trust in the tagging system, potentially leading employees to ignore warnings altogether. Additionally, implementers must consider the computational overhead of real-time detection and the potential for attackers to adapt their strategies to evade detection. Future work should explore adversarial testing methodologies to continuously improve tagging systems against evolving LLM-based threats from both known and previously unseen models. Behavioral anomaly detection systems could monitor communication patterns, flagging deviations from established sender behaviors such as unusual timing, atypical requests, or shifts in writing style \cite{wosah2024analysing}. Organizations might also implement adaptive authentication protocols that escalate verification requirements when suspicious behaviors are detected \cite{unsel2023risk}.  These solutions should complement, rather than replace, user education focused specifically on identifying sophisticated LLM-generated phishing attempts. While email security solutions with advanced detection capabilities represent a significant initial investment, this cost must be contextualized against the potentially greater financial impact of security breaches. The long-term benefits of robust email security, including breach prevention and risk mitigation, can outweigh the upfront expenditure.

The threat landscape will likely evolve significantly as LLMs advance beyond current capabilities. Future iterations of models OpenAI's GPT, Anthropic's Claude, Google's Gemini, and even open-source models such as Meta's LLAMA models may be able to generate even more convincing phishing content with improved contextual understanding and reduced linguistic artifacts. Of particular concern is the emergence of multi-modal phishing attacks that combine text with deepfake audio or video elements \cite{schmitt2024digital}. For instance, a phishing email might include a fabricated video conference recording or voicemail from a supervisor, substantially increasing the attack's credibility. These multi-modal approaches could exploit established trust relationships while circumventing traditional text-based detection methods. 

\subsection{Overall discussion and findings}

Discussions with the cyber operations team have yielded several key takeaways from the findings of the experiments. Firstly, there is a pressing need to raise awareness about the sophistication of phishing emails. Particularly with the help of LLMs, such emails may not always contain typos or obvious red flags; they can be well-crafted and may even appear to come from a trusted supervisor. The training and awareness programs need to emphasize that any account, particularly those in higher roles, can be compromised and pose significant risks to the organization. The impact of campaigns targeting high-ranking roles could be considerable, and thus, any security incidents involving VIPs or high-level accounts should be treated with utmost urgency \cite{yadav2015review, pienta2020protecting, gusev2022domestic}. Given that the experiments showed that the supervisor to direct report phishing attacks were highly successful, the rationale is to quickly address the heightened risk of the attack spreading more effectively throughout the organization since attacks can move laterally within the organization \cite{bowman2020detecting}. Such attacks can lead to extensive incident remediation efforts and substantial financial implications. Moreover, the ripple effect of these phishing emails, spreading both internally and externally, poses a serious threat to the organization's reputation. It elevates the risk of a data breach and can potentially harm long-standing relationships. Prioritizing incidents involving supervisory accounts acknowledges the amplified risk these positions carry and is a strategic move to contain and mitigate the wider organizational impact of such phishing attacks.

The after action review conducted by the cyber operations team yielded crucial insights into the employee preparedness in identifying phishing emails, particularly those generated by LLMs. Given the sophistication of LLM-generated emails, which often lack grammatical errors and appear highly credible, it's imperative to enhance the security awareness training program. This updated training should focus on educating employees to discern subtle cues, such as the overall context of the email and the nature of the request. Alongside recognizing these nuances, it is equally important for employees to be alert to common phishing red flags like the urgency of requests and the origin of URLs or attachments. Table \ref{tab:role_interaction} reveals that Student Roles were the most vulnerable job role in this lateral phishing exercise. This finding corroborates the observations made by the cyber operations team during real-world phishing attacks faced by the organization. The data suggests that focused phishing awareness training may be necessary for this more vulnerable population to effectively mitigate the risks associated with targeted phishing attempts.

Beyond identifying vulnerable populations, temporal factors may also significantly influence phishing susceptibility. Our conversations with the cyber operations team suggest that vulnerability likely fluctuates with organizational cycles and workload patterns. Specifically, the beginning of academic semesters represents a period of heightened vulnerability when many employees haven't completed recent security training and when students are particularly susceptible to job-related phishing schemes. Other temporal risk factors include tax season, institutional financial deadlines, and periods following the publication of news releases featuring organizational members. These individuals often become immediate targets for personalized phishing attempts. While our current study design doesn't capture these temporal variations, understanding these cyclical patterns could inform more strategic deployment of security resources and training refreshers during predicted high-risk periods.

Based on our findings, we propose several targeted enhancements to existing security awareness programs. Our data indicates that students workers are particularly vulnerable to phishing attempts. To address this vulnerability, we recommend a multi-faceted approach that leverages established social networks within the organization to disseminate educational content specifically tailored to at-risk populations. This approach would be complemented by a two-phase training structure consisting of comprehensive annual training followed by targeted refresher sessions that reinforce key concepts throughout the year. The establishment of student ambassador programs would further strengthen these initiatives by engaging representatives from the student population in both the ideation and feedback phases of security awareness development. The content of these enhanced training programs should emphasize that sophisticated phishing attempts may be professionally written and indistinguishable from legitimate communications from a stylistic perspective. This represents a significant shift from traditional training that often focuses on identifying grammatical errors or obvious inconsistencies. Scenario-based exercises simulating LLM-generated phishing attempts would train employees to assess contextual factors rather than superficial textual elements.

To measure the effectiveness of these enhanced training initiatives, we propose establishing a target phishing link click rate of less than 5\% across the organization, with specific monitoring mechanisms for vulnerable populations such as students. Year-over-year comparative analysis using standardized phishing simulations (consistent topic, urgency and sender) would provide quantitative metrics of improvement, while qualitative feedback collected through student focus groups would offer insights into the efficacy of training and knowledge retention. While some of these trainings can largely be incorporated within existing training frameworks without significant additional expenditure, developing and delivering specialized training for specific subgroups may require additional resources. However, the targeted approach is justified by the disproportionate vulnerability demonstrated by certain populations and the potential organizational impact of successful phishing attacks involving these groups.

\subsection{Limitations and Future Work}

While our current study provides valuable insights into the effectiveness of LLM-generated phishing emails compared to human-crafted ones, we acknowledge several limitations that present opportunities for future research. First, our short-term experimental design could be expanded into longitudinal studies to assess the sustained effectiveness of countermeasures such as email tagging systems and specialized awareness training programs. Such long-term studies would help determine whether the initial awareness and vigilance developed through our proposed trainings persist over time or require periodic reinforcement. Second, as AI technologies evolve, the threat landscape extends beyond text-based phishing. Future work could examine multimodal phishing attacks that combine LLM-generated text with other AI-generated content such as deepfakes, synthetic voice (voice phishing or "vishing"), or AI-generated images. These multi-vector attacks could potentially bypass traditional text-based detection methods and exploit different cognitive vulnerabilities in users. Third, our study measured credential entry as a binary outcome (entered vs. not entered), which provides clear effectiveness metrics but lacks granularity in understanding user hesitation behaviors. Future research could incorporate behavioral metrics such as session duration before submission, partial form completion patterns, and abandonment rates to better understand the decision-making process that occurs between receiving a phishing email and potential credential submission.

Our study's focus on a higher education institution presents both advantages and limitations for generalizability to other types of organizations. Universities operate within unique organizational structures characterized by highly decentralized administration \cite{manning2017organizational} similar to governmental organizations but distinct from many corporate environments, frequent personnel turnover particularly among student workers, and diverse demographic compositions across departments. Higher education institutions also face targeted phishing attempts due to the abundance of publicly available information about employees, making them attractive targets for social engineering \cite{casagrande2023alpha}. While these characteristics provided a rich environment for our study, they also necessitate caution when extrapolating our findings to other organizational contexts. Future research could replicate our methodology across different sectors including corporate environments, healthcare institutions, and government agencies to establish the generalizability of our findings and develop context-specific defensive strategies that account for varying organizational structures, personnel dynamics, and security practices.

\section{Conclusion}
\label{conclusion}

In this study, we have presented a comparative analysis of the effectiveness of lateral phishing emails generated by LLMs and those crafted by human communications professionals within a large educational organization. Our findings underscore the pressing threat posed by LLM-generated phishing emails, as they have proven to be as effective as, and in some cases even more effective than, their human-crafted counterparts. The experiments conducted across roughly 9,000 employees at the university revealed that these phishing emails can exploit trust relationships and power dynamics within the organization.
Our analysis of click-through rates across different job roles and department affiliations identified potential patterns of vulnerability to lateral phishing attempts. These initial findings can help guide the development of awareness campaigns and the strategic allocation of resources to bolster defenses where they are needed most.
The questionnaire administered to employees who entered data in the simulated phishing emails provided valuable insights into the underlying reasons and motivations for engaging with phishing content. The sender's identity, the perceived relevance of the email, and the sense of urgency were identified as key factors influencing recipients' decisions to interact with the emails. This highlights the need for comprehensive phishing awareness training that addresses lateral phishing and the exploitation of compromised internal accounts.

While the capability of LLMs to generate human-like text may seem intuitive, our study provides the first empirical quantification of this threat in a large-scale organizational context, offering insights for security practitioners on specific vulnerability patterns and countermeasures. This study contributes to the growing body of research on the emerging threat of LLM-generated phishing attacks and emphasizes the critical need for organizations to adapt their cybersecurity strategies accordingly. The findings underscore the importance of investing in robust employee training and awareness programs that specifically address the unique characteristics of AI-powered phishing attempts. Additionally, the insights gathered from this study can inform the development of advanced technological defenses, such as email tagging for LLM-generated content, to augment users' ability to identify and avoid these threats.
As the landscape of cybersecurity threats continues to evolve, organizations must remain vigilant and proactive in mitigating the risks posed by LLM-generated phishing attacks.

\bibliographystyle{unsrt}
\bibliography{bibliography_2}

\begin{IEEEbiography}[{\includegraphics[width=1in,height=1.25in,clip,keepaspectratio]{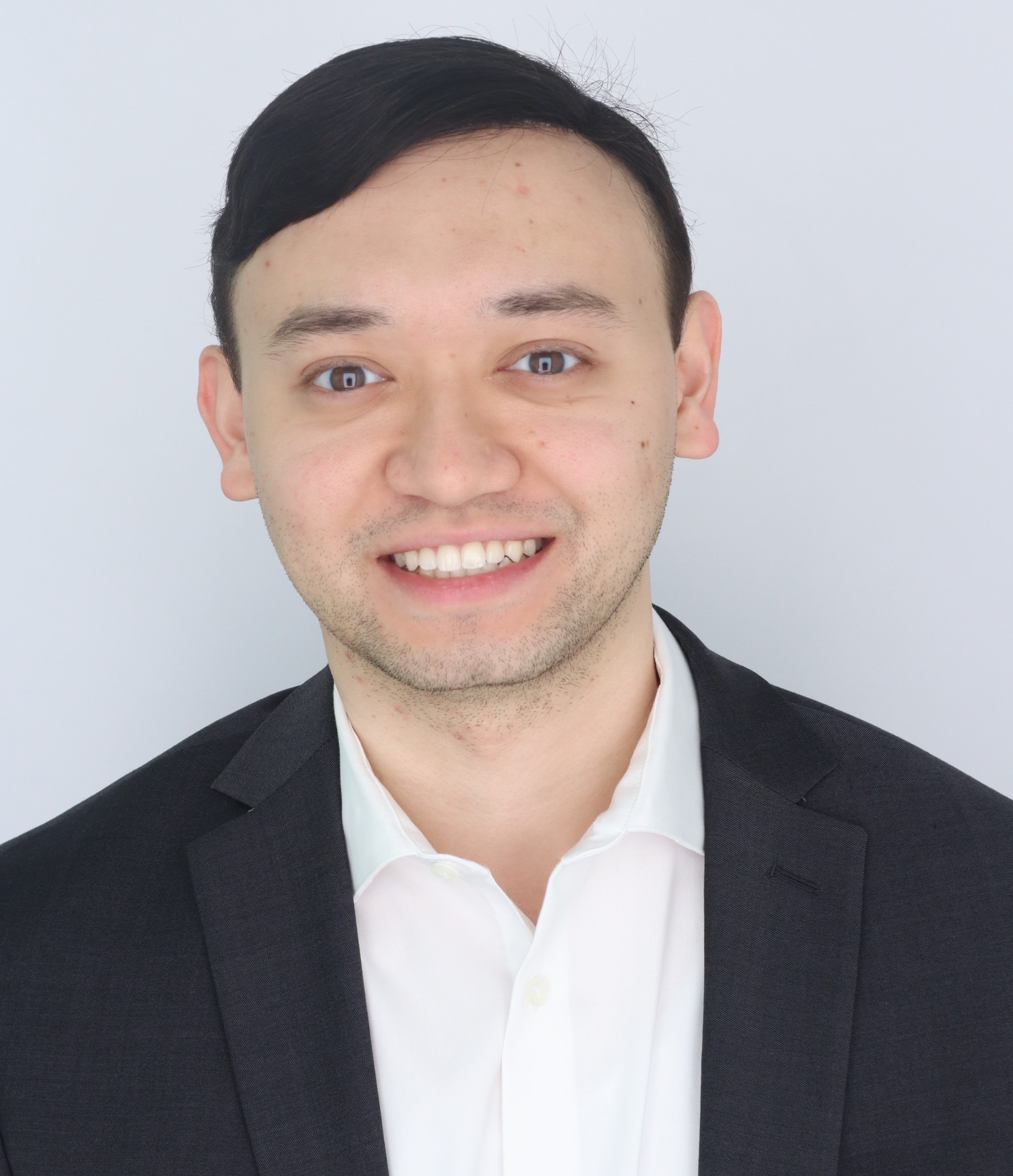}}]{Mazal Bethany} is a Ph.D. Candidate in the Department of Information Systems and Cyber Security at the University of Texas at San Antonio (UTSA), San Antonio, TX, USA. Mazal received a Bachelor of Arts degree in Economics from the University of Texas at Austin in 2020. He is currently a Graduate Research Assistant working in the Secure AI and Autonomy lab at UTSA. His research interests lie at the intersection of artificial intelligence and cybersecurity, focusing on leveraging AI for security applications, investigating vulnerabilities in AI systems, and exploring broader AI applications for business and society.
\end{IEEEbiography}

\begin{IEEEbiography}[{\includegraphics[width=1in,height=1.25in,clip,keepaspectratio]{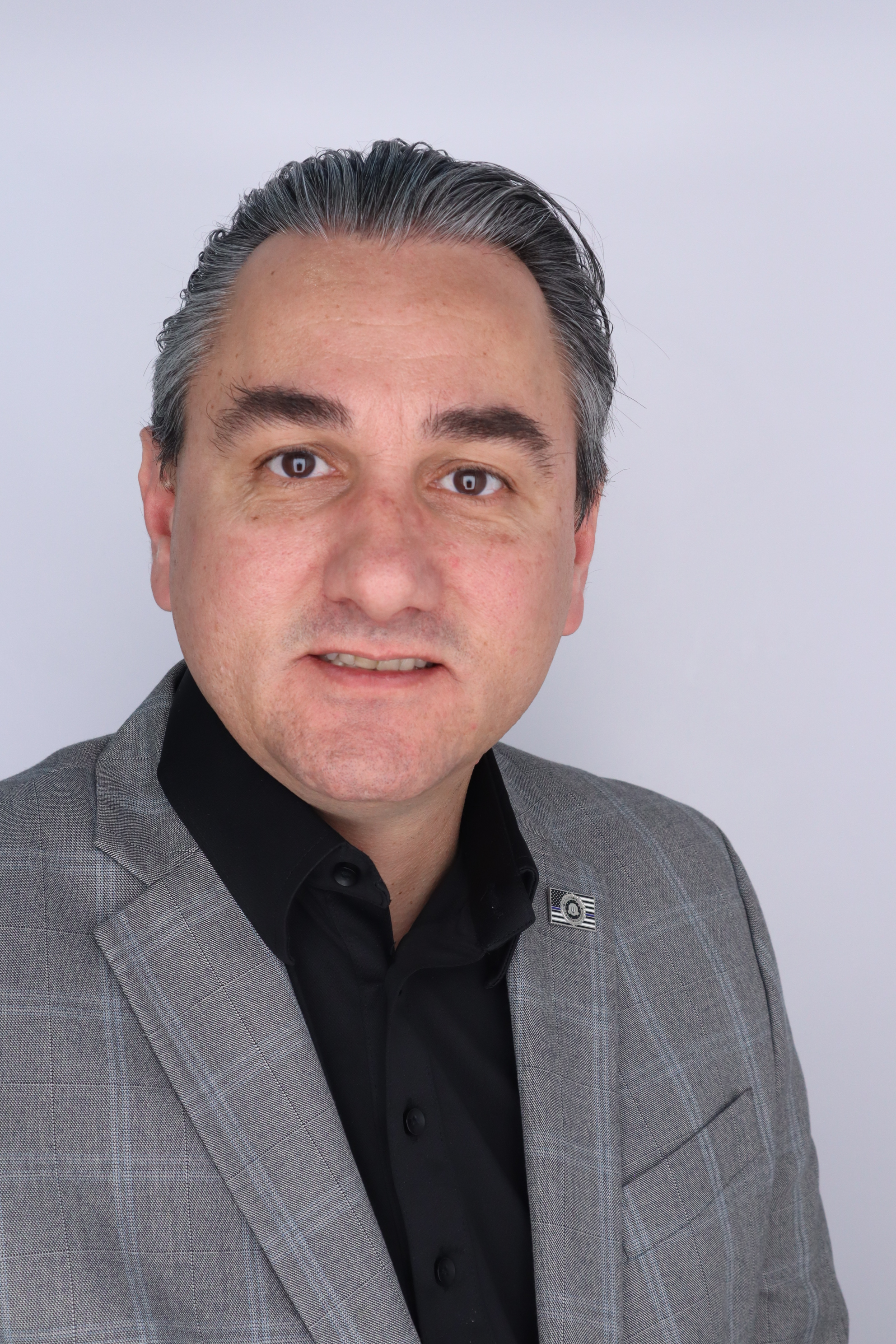}}]{Athanasios Galiopoulos} Chief Technology Officer and Deputy CIO at the University of Texas at San Antonio, is a dual citizen of the US and Greece with over 20 years of experience in strategic digital transformation and operational optimization. He has led significant initiatives, including the development of the Advanced Cyber Infrastructure Research Platform, contributing to UTSA’s R1 research status, and secured a \$15M grant for high-performance computing. With expertise in cyber AI, machine learning, and educational technology, Nassos has achieved over \$25M in cost savings and driven improvements in customer satisfaction and operational efficiency. Recognized with the FutureEdge 50 Award and featured in "Agents of Transformation," his leadership philosophy is centered on fostering high-performing teams.
\end{IEEEbiography}

\begin{IEEEbiography}[{\includegraphics[width=1in,height=1.25in,clip,keepaspectratio]{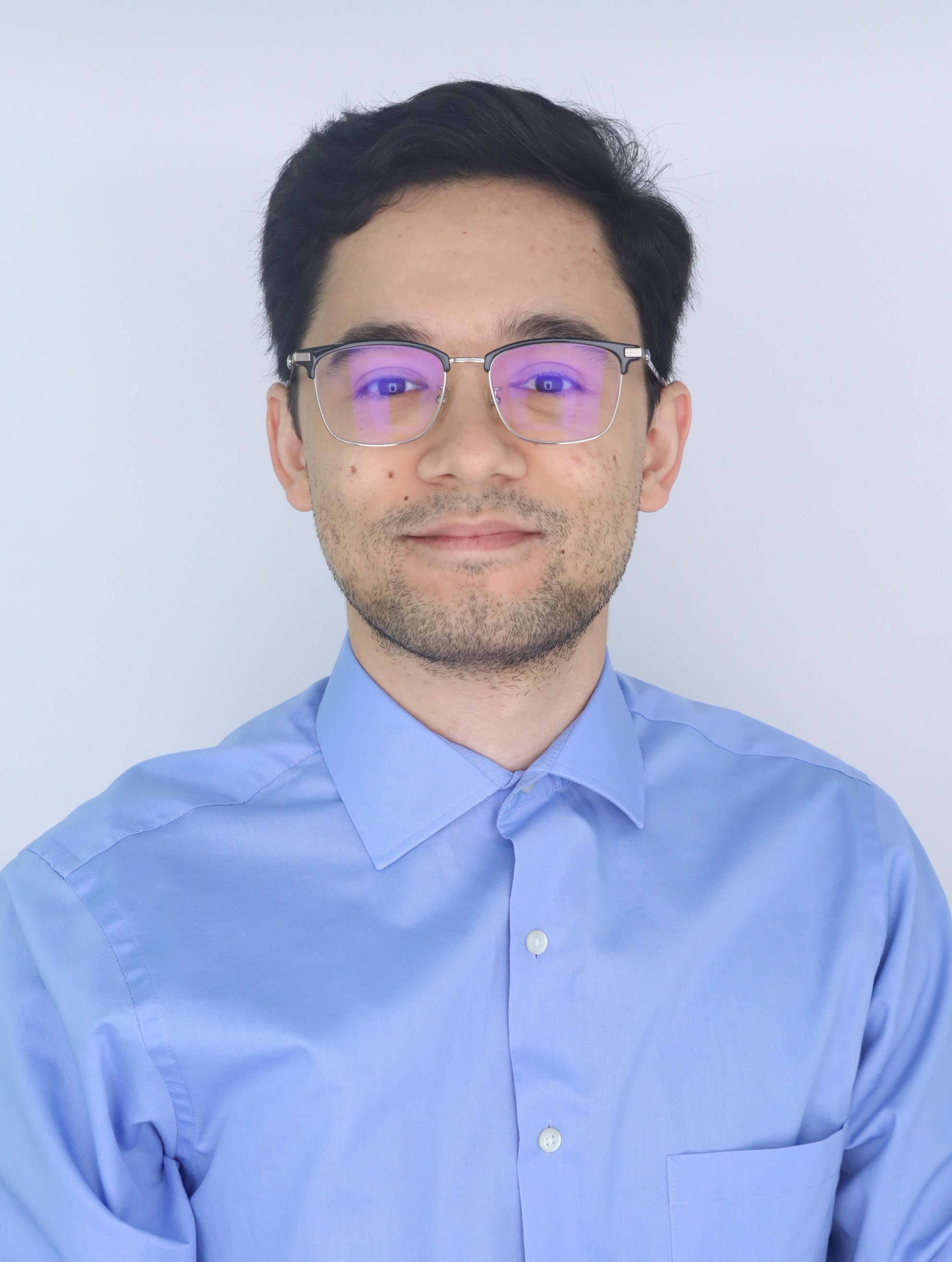}}]{Emet Bethany} is a Ph.D. student in Computer Science at the University of Texas at San Antonio. He received his M.S. in Statistics from Texas State University in 2023 and his B.S. in Applied Mathematics from the same institution in 2021. He currently also works as an AI Research Scientist at Peraton Labs, where he applies his expertise in artificial intelligence and data analysis. His research focuses on the safety and security of Large Language Models, and improving the reasoning capabilities of AI models.
\end{IEEEbiography}

\begin{IEEEbiography}[{\includegraphics[width=1in,height=1.25in,clip,keepaspectratio]{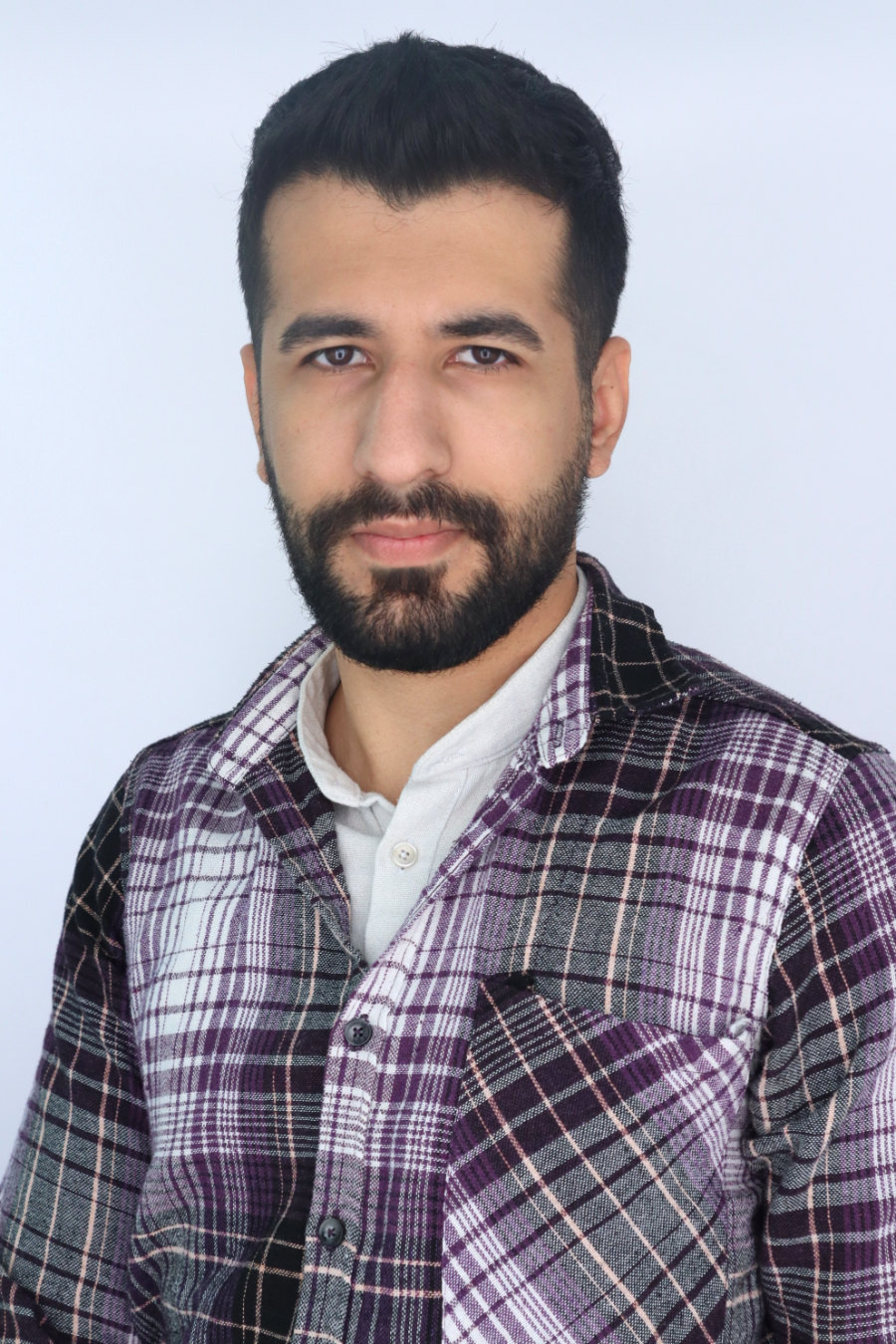}}]{Mohammad Bahrami Karkevandi}  is a Ph.D. student in the Computer Science Department at the University of Texas at San Antonio, conducting research within the Secure AI and Autonomy Lab. He holds a bachelor's degree in computer engineering from Isfahan University of Technology (2023). His research focuses on safety and privacy in large language models, model editing, unlearning and knowledge erasure, red teaming and defenses for language models, and ensuring faithfulness in AI-generated outputs.
\end{IEEEbiography}

\begin{IEEEbiography}[{\includegraphics[width=1in,height=1.25in,clip,keepaspectratio]{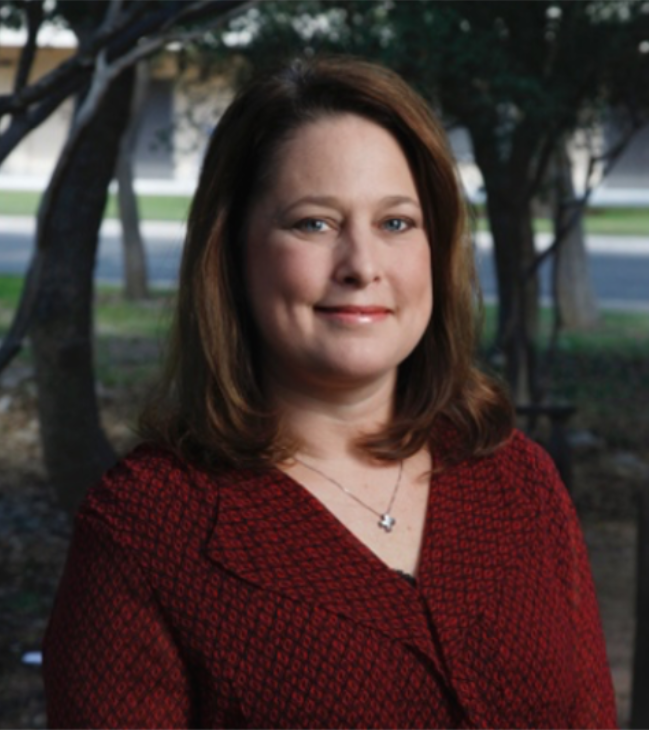}}]{Nicole Beebe} is the Melvin Lachman Endowed Chair and a Professor of Cyber Security, at the University of Texas at San Antonio (UTSA). She is currently the Assistant Vice President for Faculty Research Development, in the Office of Research at UTSA. She previously served as the interim Associate Vice President for Research Partnerships \& Strategy and the department chairperson for the UTSA Department of Information Systems \& Cyber Security. Dr. Beebe received her Ph.D. in Business Administration - Information Technology from UTSA, an M.S. degree in Criminal Justice from Georgia State University, and a B.S. degree in Electrical Engineering from Michigan Technological University. She has over twenty-five years of experience in cybersecurity and digital forensics, in both the commercial and government sectors. Her research interests relate to cybersecurity, cyber analytics, and digital forensics, with applications to insider threat detection and analysis, IoT security and forensics, and cyber threat hunting.
\end{IEEEbiography}

\begin{IEEEbiography}[{\includegraphics[width=1in,height=1.25in,clip,keepaspectratio]{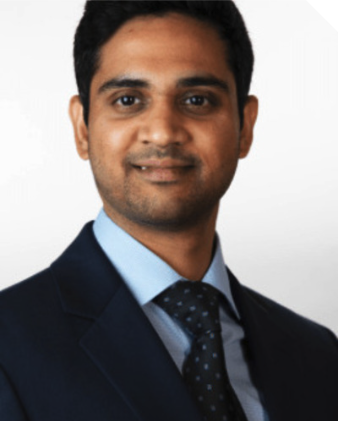}}]{Nishant Vishwamitra} received his Ph.D. in Computer Science and Engineering in 2022 from the State University of New York at Buffalo. He currently holds the position of assistant professor in the Department of Information Systems and Cyber Security at the University of Texas at San Antonio. His research is focused on addressing diverse, emerging cybersecurity threats such as online  abuse, cyberbullying, cyber harassment, online offensiveness, attacks and defenses of AI/ML models, image privacy, and crowdsourcing. Dr. Vishwamitra also collaborates with  researchers in other domains such as Social Science, Sociology, and Communication. He is on the Artifacts Evaluation Committee of the Annual Computer Security Applications Conference (ACSAC), 2021, and the review committee of the IEEE International Workshop Big Data Security and Services (Big Data Service), 2018–2020. He is  a reviewer for several renowned cybersecurity and AI conferences and journals. 
\end{IEEEbiography}

\begin{IEEEbiography}[{\includegraphics[width=1in,height=1.25in,clip,keepaspectratio]{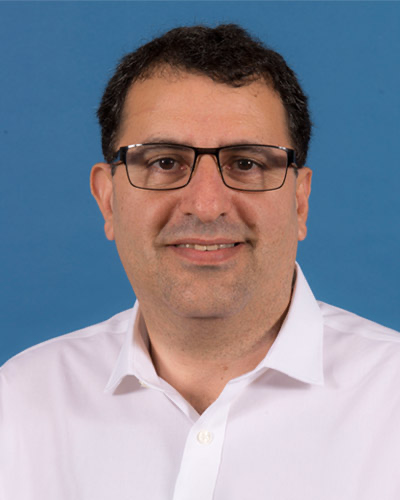}}]{Peyman Najafirad (Paul Rad)} is the Associate Director of Research at the School of Data Science and an Associate Professor with joint appointments in the Computer Science, Information Systems, and Cyber Security departments at the University of Texas at San Antonio (UTSA). As a senior member of the National Academy of Inventors, Dr. Rad is a leading authority on trustworthy artificial intelligence and reasoning for decision-making under uncertainty. His research focuses on developing safe and secure AI systems, with particular emphasis on large language models, the use of generative AI to enhance digital content integrity, and physics-informed AI for digital twins. Dr. Rad is also the Founder and Chair of Safe AI, a startup dedicated to advancing secure and safe AI solutions for enterprises. His pioneering work has led to numerous patents, top-tier publications, and significant contributions to shaping national AI standards through his collaboration with NIST.
\end{IEEEbiography}

\EOD

\end{document}